\documentstyle[12pt]{article}

\def\journal{\topmargin .3in    \oddsidemargin .5in
        \headheight 0pt \headsep 0pt
        \textwidth 5.625in % 1.2 preprint size  %6.5in
        \textheight 8.25in % 1.2 preprint size 9in
        \marginparwidth 1.5in
        \parindent 2em
        \parskip .5ex plus .1ex         \jot = 1.5ex}
\journal

\catcode`\@=11
\def\marginnote#1{}

\begin{document}
\begin{titlepage}
\pagestyle{empty}
\begin{center}
\today     \hfill    hep-ph/9301294\\

\vskip .25in

{\large \bf A More-Effective Potential}
\footnote{This work was supported by the Director, Office of Energy 
Research, Office of High-Energy and Nuclear Physics, Division of 
High-Energy Physics of the U.S. Department of Energy under contracts 
DE-AC03-76SF00098 and DE-FG04-84ER40166.}

\vskip .2in

Kevin Cahill \\[.15in]

{\em  Department of Physics and Astronomy,
      University of New Mexico\\
      Albuquerque, New Mexico 87131-1156
\footnote{Permanent address,  e-mail: kevin@cahill.unm.edu}\\
      Theoretical Physics Group,
      Lawrence Berkeley Laboratory\\
      Berkeley, California 94720}
\end{center}

\vskip .2in

\begin{abstract}
In theories with spontaneous symmetry breaking,
the loop expansion of the effective potential 
is awkward. 
In such theories, the exact effective potential $V(\phi_c,T)$ is 
real and convex (as a function of the classical field $\phi_c$),
but its perturbative series can be complex with  
a real part that is concave.
These flaws limit the utility of the effective potential,
particularly in studies of the early universe.
A generalization of the effective potential
is available that is real, that has
no obvious convexity problems, and that
can be computed in perturbation theory.
For the theory with classical potential 
$V(\phi) = (\lambda/4)(\phi^2 - \sigma^2)^2$,
this more-effective potential closely tracks
the usual effective potential where the latter is real
$|\phi_c| \geq \sigma/\sqrt{3}$
and naturally extends it to $\phi_c = 0$,
revealing that the critical temperature
at the one-loop level
runs from $T_C \approx 1.81 \sigma$ for $\lambda = 0.1$
to $T_C \approx 1.74 \sigma$ for $\lambda = 1$.
\end{abstract}
\end{titlepage}
%
%THIS PAGE (PAGE ii) CONTAINS THE LBL DISCLAIMER
%TEXT SHOULD BEGIN ON NEXT PAGE (PAGE 1)
%\renewcommand{\thepage}{\roman{page}}
%\setcounter{page}{2}
%\mbox{ }
%\vskip 1in
%{\bf Disclaimer}
%\vskip .2in
%This document was prepared as an account of work sponsored by the United
%States Government.  Neither the United States Government nor any agency
%thereof, nor the Regents of the University of California, nor any of their
%employees, makes any warranty, express or implied, or assumes any legal
%liability or responsibility for the accuracy, completeness, or usefulness
%of any information, apparatus, product, or process disclosed, or represents
%that its use would not infringe privately owned rights.  Reference herein
%to any specific commercial-products process, or service by its trade name,
%trademark, manufacturer, or otherwise, does not necessarily constitute or
%imply its endorsement, recommendation, or favoring by the United States
%Government or any agency thereof, or the Regents of the University of
%California.  The views and opinions of authors expressed herein do not
%necessarily state or reflect those of the United States Government or any
%agency thereof or of the Regents of the University of California and shall
%not be used for advertising or product-endorsement purposes.
%\vskip 2in
%{\it Lawrence Berkeley Laboratory is an equal-opportunity employer.}
%
\newpage
\renewcommand{\thepage}{\arabic{page}}
\setcounter{page}{1}
%THIS IS PAGE 1 (INSERT TEXT OF REPORT HERE)
\goodbreak
     
\section{\bf Introduction}
\par
The effective potential was introduced
by Heisenberg and Euler~\cite{Heis} and by Schwinger~\cite{Schw}.
Goldstone, Salam, and Weinberg~\cite{Gold 62}
and Jona-Lasinio~\cite{Jona 64}
developed the effective potential and applied it
to the problem of symmetry breaking~\cite{Gold 61}.
Coleman and E.~Weinberg used it to
show that radiative corrections could break symmetries~\cite{Cole 73}.
Linde~\cite{Lind 76} and S.~Weinberg~\cite{Wein 76}
later used it to obtain a lower bound on the mass
of the Higgs boson.
West and others have used it
to study the breaking of supersymmetry~\cite{West 76}.
\par
The possibility that broken symmetries 
might be restored at high temperatures
was raised by Kirzhnits and Linde~\cite{Lind 72}
and confirmed by them~\cite{KLind 74},
by Dolan and Jackiw~\cite{Dola 74}
and by S.~Weinberg~\cite{Wein 74}
who with Bernard~\cite{Bern 74}
introduced and developed the finite-temperature effective potential.
Much current work on the early universe 
is based upon the finite-temperature effective potential~\cite{Kolb 90}.
\par
Although the effective potential has a long
history of successful applications to particle physics,
it does not seem to be well suited
to theories that exhibit spontaneous symmetry breaking.
In such theories the loop expansion of the effective potential 
can be awkward~\cite{loopsfail}.
While the exact effective potential is both real and convex
(as a function of the mean values $\phi_c$
of the scalar fields)~\cite{real&convex},
its perturbative series 
in theories with spontaneously broken symmetry
displays neither property.
In the example provided by 
by the symmetry-breaking classical potential
$V(\phi) = (\lambda/4)(\phi^2-\sigma^2)^2$, 
the loop expansion 
of the finite-temperature effective potential
is complex at all temperatures $T$
for $|\phi_c| < \sigma/\sqrt{3}$,
and its real part is concave at low temperatures
and small $\lambda$ for a similar range of $|\phi_c|$. 
In such theories 
the accuracy of the one-loop effective potential does not
extend down to the small values of $|\phi_c|$
that are of interest in studies of the early universe. 
\par
Of these two issues, convexity and complexity,
it is the complexity that is the more serious.
The convexity of the exact effective potential 
(with an ultraviolet regulator in place) suggests 
that the effective potential may not be the ideal tool
for studying systems with spontaneous symmetry breaking
in quasi-equilibrium in finite regions of spacetime.  
The concavity of its renormalized loop expansion 
in turn suggests that the loop expansion 
may be an uncertain approximation to the effective potential
for such systems.
But the critical defect of the effective potential 
is the complexity of its loop expansion.
For where the effective potential is complex, 
it is ambiguous as an approximation to a free-energy density ---
although it may be interpreted as a decay rate~\cite{EW}.
\par
Because of these limitations of the perturbative effective potential,
some physicists have turned to nonperturbative techniques.
Chang~\cite{Chang} has invented a variational method
called the gaussian effective potential,
which Barnes and Ghandour~\cite{Barnes} 
and Stevenson~\cite{Stevenson} have developed.
Fukuda and Kyriakopoulos~\cite{Fukuda}
have introduced a version of the effective potential
that is well suited to lattice computations;
O'Raifeartaigh, Wipf, and Yoneyama~\cite{O'Raifeartaigh} 
have analyzed this potential.
Ringwald and Wetterich~\cite{Ringwald} 
have suggested the use of block-spin techniques.
\par
The goal of the present paper is to generalize
the effective potential so that it can be applied
simply and perturbatively to theories with
spontaneous symmetry breaking.
The usual effective potential
is the Legendre transform of the Helmholtz free-energy density
for the modified hamiltonian $H + \int j\phi \, d^3x$
in which $j$ is an external source.
For theories with potentials of positive curvature,
$V''(\phi) \ge 0$,
the probe $j \phi$ is optimal. 
But for theories in which $V''(\phi)$ 
takes on negative values,
as it must when $V(\phi)$ has two minima,
it is argued that
the linear probe $j \phi$ be generalized to a 
quadratic polynomial $j P(\phi)$.
This advice has been given 
in the past with varying degrees of obliqueness
by Cornwall, Jackiw, and Tomboulis~\cite{Corn 74},
by Hawking and Moss~\cite{Hawk}, and
by Lawrie~\cite{Lawrie}; 
but it has not been followed.
When discussing theories with spontaneous
symmetry breaking,
most physicists either ignore the
complexity of the usual effective potential~\cite{Kolb 90}
or work in a region of parameter space
in which scalar loops can be ignored~\cite{Hall}.
\par
In what follows I shall discuss
the case of a single scalar field $\phi$
interacting with an arbitrary renormalizable potential $V(\phi)$.
If the curvature $V''(\phi)$ of the potential is positive,
then the usual effective potential with a linear probe $j \phi$
is adequate.
But if curvature $V''(\phi)$ of the potential is negative
for some values of the field $\phi$,
then a quadratic polynomial $j P(\phi)$ should be used. 
If the potential $V(\phi)$ of indefinite curvature
has a single minimum at $\phi_1$,
then a suitable probe is 
$ P(\phi) = \left( \phi - \phi_1 \right)^2/2 $.
If the potential $V(\phi)$ of indefinite curvature
has two minima
separated by a local maximum at $\phi_0$,
then I suggest using $P(\phi) = ( \phi - \phi_0)^2/2$.
Such quadratic probes $ j P(\phi) $
lead to more-effective potentials that
possess real loop expansions
and have no obvious convexity problems.
\par
For the classical potential $V(\phi) = (\lambda/4)
(\phi^2 - \sigma^2)^2$, which has a local maximum
at $ \phi_0 =  0 $, 
the appropriate probe is $P(\phi) = \phi^2/2$.
The resulting effective potential 
closely tracks the usual effective potential where the latter
is real and naturally extends it down to $\phi_c=0$. 
The reflection symmetry of the action is restored
to the vacuum at a critical temperature $T_C$
which runs from $T_C \approx 1.81 \, \sigma$ for $\lambda = 0.1$
to $T_C \approx 1.74 \, \sigma$ for $\lambda = 1$.
The first few terms of the high-temperature
expansion of the one-loop effective potential 
$V_1(\phi_c,T;P)$ with probe $P(\phi) = \phi^2/2$ are
\begin{equation}
V_1(\phi_c,T;P) = 
- {\pi^2 \over 90} T^4
+ {\lambda \over 24} \left( 3 \phi_c^2 - \sigma^2 \right) T^2
- {\lambda^{3/2} \over 12 \sqrt{2} \, \pi}
\left( 7|\phi_c|^3 - 3\sigma^2 |\phi_c| \right) T.
\label{ht mep}
\end{equation}
The corresponding terms of the usual effective potential are
\begin{equation}
V_1(\phi_c,T) =  - {\pi^2 \over 90} T^4
+ {\lambda \over 24} \left( 3\phi_c^2 - \sigma^2 \right) T^2
- { \lambda^{3/2} \over 12 \pi }
\left( 3 \phi_c^2 - \sigma^2 \right)^{3/2} T.
\label{ht ep}
\end{equation}
These two expansions possess the same two leading terms,
but they differ in the third term,
which is imaginary for $|\phi_c| < \sigma/3$
in the case of the usual effective potential.
In the expansion of the more-effective potential
$V_1(\phi_c,T;P)$, the term
$ \lambda^{3/2} \sigma^2 |\phi_c| T $
occurs with a positive coefficient and
may have astrophysical implications\cite{Tom and me}.  
\par
The traditional effective potential is
discussed in Sections II--V
in a pedagogical manner
inspired by Weinberg~\cite{Wein 74}.
The problems that can arise in
theories that exhibit spontaneous
symmetry breaking are illustrated
in Sec.~VI\null.
The more-effective potential
is introduced in Sec.~VII\null.
The meaning of generalized effective
potentials is discussed in Sec.~VIII\null.
In Sec.~IX the computation
of the generalized effective potential
is discussed for the case of a quartic
polynomial $V(\phi)$ with two minima.
This computation is carried out in
detail for the potential $V(\phi) = (\lambda/4) 
(\phi^2 - \sigma^2)^2$ in Sec.~X.
\goodbreak

\section{\bf The Partition Function of a Free Field}
\par
One of the clearest descriptions of the finite-temperature
effective potential is S.~Weinberg's operator formulation~\cite{Wein 74}.
Stripped of fermions and gauge fields
and reduced to a single scalar field,
it will serve as the basis for the introductory sections
of this paper.
\par
A free, real scalar field $\phi$ of mass $m$ 
is described by the hamiltonian 
\begin{eqnarray}
H & = & \int \! {\textstyle {1\over2}} \left( \pi^2 + 
(\nabla \phi)^2 + m^2 \phi^2\right) d^3x \nonumber \\
& = & \sum_{k} \omega_k
\left[ a^\dagger(k) a(k) + {\textstyle {1\over2}}\right]
\end{eqnarray}
where $\omega_k = \sqrt{k^2 + m^2}$.
At temperature $T$ its partition function is
\begin{equation}
Z(\beta) = \mbox{\rm Tr\,} e^{-\beta H}
\end{equation}
where $\beta = 1/T$ is the inverse temperature. 
By inserting a complete set of energy eigenstates,
one may find for the partition function $Z(\beta)$
the expression
\begin{eqnarray}
Z(\beta) & = & \prod_k e^{-{\beta \omega_k \over 2}}
\sum_{n_k=0}^\infty e^{-\beta n_k \omega_k } \nonumber \\
& = & \prod_k {e^{-{\beta \omega_k \over 2}}
\over \left(1 - e^{-\beta \omega_k }\right)}  \label{Z =}
\end{eqnarray}
which is simpler as a logarithm,
\begin{eqnarray}
-\log Z(\beta) & = & \sum_{k}
{\beta \omega_k \over 2}
+ \log\left(1 - e^{-\beta \omega_k }\right) \nonumber \\
\qquad & = & \int\! {L^3 d^3k \over (2\pi)^3}
\left[{\beta \omega_k \over 2} 
+ \log\left(1 - e^{-\beta \omega_k }\right)
\right] \label{logZ =}
\end{eqnarray}
where $L^3$ is the volume of quantization. 
\goodbreak

\section{\bf The Effective Potential} 
\par
For a scalar field $\phi$ described by a hamiltonian $H$
perturbed by an external current $j$,
the Helmholtz free-energy density $A(j,T)$ 
is defined by the relation
\begin{equation}
e^{-\beta L^3 A(j,T)} = 
\mbox{\rm Tr\,} e^{-\beta \left(H + j\!\int\!  \phi(x)\,d^3x\right)}.
\label{Adef}
\end{equation} 
The free energy $A(j,T)$ is therefore
proportional to the logarithm
of the partition function $Z(\beta, j)$ 
\begin{equation}
A(j,T) = - {\log Z(\beta, j)\over \beta L^3}
\label{A = logZ}
\end{equation}
for the system described by the perturbed
hamiltonian $H + j\int \phi(x) d^3x$.
\par
The mean value of the field $\phi$ is a function
of the current $j$
\begin{equation}
\phi_c(j,T) \equiv \langle \phi \rangle_j = {\mbox{\rm Tr\,} \phi(x) 
e^{-\beta \left(H + j\int\!\! \phi(x) d^3x \right)} \over 
 \mbox{\rm Tr\,} e^{-\beta \left(H + j\int\!\! \phi(x) d^3x\right)}} 
\label{phi_c def}
\end{equation}
and is a derivative of the Helmholtz potential
\begin{equation}
\phi_c(j,T) = {\partial A(j,T) \over \partial j}.   \label{phisdA}
\end{equation}
The finite-temperature effective potential $V(\phi_c,T)$
is defined [10--12]
as a Legendre transform of the Helmholtz potential
\begin{equation}
V(\phi_c,T) \equiv A(j,T) - j \, {\partial A(j,T) \over \partial j} 
= A(j,T) - j \phi_c                                \label{Vdef}
\end{equation}
expressed as a function of the ``classical field'' $\phi_c$
\begin{equation}
V(\phi_c,T) = A(j(\phi_c,T),T) - j(\phi_c,T)\phi_c
\end{equation}
rather than of the current $j$.
The effective potential
may be thought of as a Gibbs free-energy density.
It is obviously real.
\par
The utility of the effective potential
derives from its ability at its minima 
to represent the unperturbed system.
For from eqs.(\ref{phisdA}) and (\ref{Vdef}), it follows that
the derivative of the effective potential
with respect to the classical field $\phi_c$
is proportional to the external current $j$
\begin{equation}
{\partial V(\phi_c,T) \over \partial \phi_c}
= {\partial j \over \partial \phi_c}
\left( {\partial A(j,T) \over \partial j}
- \phi_c \right) - j = - j.                        \label{dV}
\end{equation}
Thus the current $j$ must vanish
at the stationary points of $V(\phi_c,T)$,
\begin{equation}
0 = {\partial V(\phi_c,T) \over \partial \phi_c}
= - j. \label{j=0}
\end{equation}
At zero temperature, the minimum value of the effective
potential is the energy density of the ground state
of the system.
\par
Since it is
through the factor $\exp(-\beta j \int \phi(x) d^3x)$
that the current $j$ influences the mean value $\phi_c$,
the relationship between the current $j$ and
the mean value $\phi_c$ is inverse.  
Thus both the derivative of the mean value $\phi_c$
with respect to the current $j$ 
and that of $j$ with respect to $\phi_c$
are negative or zero:
$\partial \phi_c / \partial j \le 0$ and
$\partial j / \partial \phi_c \le 0$. 
So by differentiating the formula (\ref{dV})
with respect to $\phi_c$,
one sees that the effective potential
has a nonnegative second derivative,
\begin{equation}
{\partial^2 V(\phi_c,T) \over \partial \phi_c^2}
= - {\partial j  \over \partial \phi_c} \ge 0.
\label{convex}
\end{equation}
The effective potential is therefore 
formally convex~\cite{real&convex}
as a function of the field~$\phi_c$.
\goodbreak

\section{\bf The Effective Potential for a Free Field}
\par
In the case of a free scalar field, 
one may implement these definitions exactly.
The unitary transformation
\begin{equation}
U(j) = e^{-i\int (j/m^2) \pi(x) d^3x}
\end{equation}
displaces the field $\phi(x)$
\begin{equation}
U^\dagger \phi(x) U = \phi(x) + {j \over m^2} \label{displ'd phi}
\end{equation}
and so relates the perturbed hamiltonian 
to the unperturbed one: 
\begin{equation}
U^\dagger H U = H + 
\int\! j \, \phi(x) \, d^3x + {{L^3 j^2 \over 2m^2}} . 
\end{equation}
Thus since traces are invariant under unitary transformations,
the Helmholtz potential $A(j,T)$ for the free field 
\begin{eqnarray}
e^{-\beta L^3 \left( A + { j^2 \over 2m^2 } \right) }
& = & \mbox{\rm Tr\,} e^{-\beta \left( H 
+ \int\! j \, \phi \, d^3\!x 
+ {L^3 j^2\over2m^2} \right)} \nonumber \\
& = & \mbox{\rm Tr\,} U^\dagger e^{-\beta H} U \nonumber \\
& = & \mbox{\rm Tr\,} e^{-\beta H} = Z(\beta)
\end{eqnarray}
is related to the logarithm (\ref{logZ =}) 
of the partition function 
of the unperturbed system by the equation
\begin{equation}
A(j,T) = -{\log Z(\beta) \over \beta L^3} - {j^2 \over 2m^2}.
\label{A=}
\end{equation}
\par
The effect of the linear perturbation $j\int\phi(x)d^3x$
is to displace the field by $j/m^2$,
as shown by eq.(\ref{displ'd phi}).
So the mean value $\phi_c$ is 
\begin{equation}
\phi_c = {\partial A(j,T) \over \partial j} = - {j\over m^2}.
\label{phi_c =}
\end{equation}
One may also evaluate $\phi_c$ directly.
Since the mean value $\langle \phi \rangle$ for the unperturbed theory
is $\phi_c(0,T) = 0$,
that of the perturbed theory is simply
\begin{eqnarray}
\phi_c(j,T) & = & 
{\mbox{\rm Tr\,} \phi(x) U^\dagger e^{-\beta H} U 
\over \mbox{\rm Tr\,} U^\dagger e^{-\beta H}U}\nonumber \\
& = &
{\mbox{\rm Tr\,} U \phi(x) U^\dagger e^{-\beta H} 
\over \mbox{\rm Tr\,} e^{-\beta H}}\nonumber \\
& = & - {j\over m^2}. \label{phi_c=}
\end{eqnarray}
\par
It follows now from the definition (\ref{Vdef}),
from eqs.(\ref{A=}--\ref{phi_c=}), and 
from the formula (\ref{logZ =}) for the 
partition function $Z(\beta)$
that the exact finite-temperature
effective potential for the free scalar field
of mass $m$ is
\begin{eqnarray}
V(\phi_c,T) & = & A(j,T) - j \phi_c \nonumber \\ 
\quad & = & {\textstyle { 1 \over 2}} m^2 \phi_c^2 
- {\log Z(\beta) \over \beta L^3} \nonumber \\  
\quad & = & {\textstyle { 1 \over 2}} m^2 \phi_c^2 
+ \int\! {d^3k \over (2\pi)^3}
\left[{\omega_k \over 2}
+ {\log\left(1 - e^{-\beta \omega_k }\right) \over \beta}
\right] 
\label{V free =}
\end{eqnarray}
with $\omega_k = \sqrt{k^2 + m^2}$.
The effective potential $V(\phi_c,T)$
is real and convex.  At its absolute minimum $\phi_c=0$, 
the external current $j = - m^2\phi_c$ vanishes.
In the limit $\beta \rightarrow \infty$,
the potential $V(\phi_c,T)$
becomes the exact zero-temperature effective potential
\begin{equation}
V(\phi_c,0) = {\textstyle { 1 \over 2}} m^2 \phi_c^2 
+ \int\!\! {d^3k \over (2\pi)^3}
{\omega_k \over 2}.
\label{V free at T=0 =}
\end{equation}
\goodbreak

\section{\bf The One-Loop Effective Potential}
\par
For a scalar field interacting
with itself through a potential $V(\phi)$,
the effect of the perturbing current $j$
is to replace $V(\phi)$ by
\begin{equation}
V_j(\phi) = V(\phi) + j \phi.
\end{equation}
The absolute minimum $\bar \phi$ of this altered potential
is a root of the equation
\begin{equation}
0 = {\partial V_j(\phi) \over \partial \phi} = 
{\partial V(\phi) \over \partial \phi} + j. 
\label{dV_j = 0}
\end{equation}
To obtain the one-loop approximation to the Helmholtz
potential, we replace the altered potential $V_j(\phi)$
in the definition (\ref{Adef}) of $A(j,T)$
by its Taylor-series expansion about the absolute
minimum $\bar \phi$:
\begin{equation}
V_j(\phi) \approx V_j(\bar \phi) 
+ {\textstyle { 1 \over 2}} 
{\partial^2 V_j(\bar \phi) \over \partial \phi^2}
(\phi - \bar \phi)^2.
\label{trunc series}
\end{equation}
To zeroth order in $\hbar$,
the minimum $\bar \phi$ is the mean value 
$\phi_c$ of the scalar field $\phi$
as defined by eq.(\ref{phi_c def}). 
The truncated series (\ref{trunc series}) 
\begin{equation}
V_j(\phi) \approx V(\bar \phi) + j\bar \phi
+ {\textstyle { 1 \over 2}}
{\partial^2 V(\bar \phi) \over \partial \phi^2}
(\phi - \bar \phi)^2
\end{equation}
describes a free scalar field of mass 
\begin{equation}
m = \sqrt{ V^{\prime\prime}(\bar \phi)} .
\end{equation}
The quantity $V^{\prime\prime}(\bar \phi)$ is positive
because $\bar \phi$ is a minimum of $V_j(\phi)$.
\par
One may now express the Helmholtz potential $A(j,T)$
in terms of the kinetic energy $K = \int {\textstyle(1/2)} \pi(x)^2 d^3x$ as
\begin{eqnarray}
e^{-\beta L^3 A(j,T)} 
& = & 
\mbox{\rm Tr\,} e^{-\beta \left(H + j\int\!\! \phi(x) d^3x\right)} \nonumber \\
& = & 
\mbox{\rm Tr\,} e^{-\beta \left(K + \int\!\! V_j(x) d^3x\right)} \nonumber \\
& \approx & e^{-\beta L^3 \left( V(\bar \phi) + j\bar \phi \right)} 
\mbox{\rm Tr\,} e^{-\beta \left[K + \int\!\! (1/2) m^2 (\phi(x) - \bar \phi)^2
d^3x\right]}.
\label{exp(-A) = Z}
\end{eqnarray}
So by using the unitary operator
\begin{equation}
U = e^{i\int \bar \phi \pi(x) d^3x}
\label{U2}
\end{equation}
which displaces the field $\phi(x)$ to
$U^\dagger \phi(x) U = \phi(x) - \bar \phi$,
one may write $A(j,T)$ approximately as
\begin{eqnarray}
e^{-\beta L^3 A(j,T)}
& \approx & e^{-\beta L^3 \left( V(\bar \phi) + j\bar \phi \right)} \:
\mbox{\rm Tr\,} [U^\dagger e^{-\beta H_0} U] \nonumber \\ 
& \approx & e^{-\beta L^3 \left( V(\bar \phi) + j\bar \phi \right)} \:
\mbox{\rm Tr\,} [ e^{-\beta H_0} ] \nonumber \\ 
& \approx & e^{-\beta L^3 \left( V(\bar \phi) + j\bar \phi \right)} Z(\beta)
\end{eqnarray}
where $Z(\beta)$ is the exact partition function (\ref{logZ =})
for the free scalar field of mass $m$.
Thus at the one-loop level,
the Helmholtz potential is
\begin{equation}
A_1(j,T) = V(\bar \phi) + j\bar \phi - {\log Z(\beta) \over \beta L^3}.
\label{A_1(j,T) =}
\end{equation}
\par
The mean value $\phi_c$ and the mean $\bar \phi$
differ only by terms of order $\hbar$,
which are due to the quantum fluctuations induced
by the kinetic energy $K$.
Specifically it follows from eqs.(\ref{phisdA}) and (\ref{dV_j = 0}) 
that this difference is
\begin{equation}
\phi_c = \bar \phi - {\partial \over \partial j} 
{\log Z(\beta) \over \beta L^3}.
\label{phi_c is}
\end{equation}
Thus by the extremal condition (\ref{dV_j = 0}),
the altered potential changes only by quantities
that are of second order in $\hbar$ as $\bar \phi$
is replaced by $\phi_c$,
\begin{equation}
V(\bar \phi) + j\bar \phi =  V(\phi_c) + j\phi_c + {\cal O} (\hbar^2).
\label{V_j stationary}
\end{equation}  
We may therefore write the Helmholtz potential
to first order in $\hbar$ 
in terms of the mean value $\phi_c$ of the field $\phi$
\begin{equation}
A_1(j,T) = V(\phi_c) + j\phi_c - {\log Z(\beta) \over \beta L^3}
\label{A_1(j,T) phi_c}
\end{equation}
in which we have also freely replaced $\bar \phi$ by $\phi_c$
in the logarithm of $Z$ which itself is of order $\hbar$.
Now by performing the Legendre transform (\ref{Vdef}), 
we find that the effective potential is
\begin{eqnarray}
V_1(\phi_c,T) & = & V(\phi_c) 
- {\log Z(\beta) \over \beta L^3}\nonumber \\
\quad & = & V(\phi_c) 
+ \int\!\! {d^3k \over (2\pi)^3}
\left[{\omega_k \over 2}
+ {\log\left(1 - e^{-\beta \omega_k }\right) \over \beta}
\right]
\label{V_1(phi_c,T) =}
\end{eqnarray}
with
\begin{equation}
\omega_k = \sqrt{ k^2 + V^{\prime\prime}(\phi_c)} , 
\label{omega = sqrt}
\end{equation}
which is the usual result.
\par
Classical potentials that induce spontaneous symmetry breaking
have second derivatives that are negative between their inflection points. 
When the second derivative $V^{\prime\prime}(\phi_c)$ is negative,
the frequency $\omega_k$ becomes complex for small enough $k$,
and the loop expansion for the effective potential fails.
\par
The preceding integral of $\omega_k$ over $d^3k$ diverges.
We may renormalize the effective potential
by interpreting the classical potential $V(\phi)$
as containing counterterms $V_{ct}(\phi)$ of order $\hbar$
that are the same form as the terms of $V(\phi)$,
apart from a constant term.
By introducing a cut-off $\Lambda$
and performing the integration,
we find for the Helmholtz potential the expression
\begin{eqnarray}
A_1(j,T) & = & V(\phi_c) + j\phi_c
+ {V''(\phi_c)^2 \over 64 \pi^2}
\left[\log{{V''(\phi_c) \over \mu^2}} + {1 \over 2} \right]
 \nonumber \\
& & \mbox{}
+ {\Lambda^4 \over 16 \pi^2} + {\Lambda^2 V''(\phi_c) \over 16 \pi^2}
+ {V''(\phi_c)^2 \over 64 \pi^2} \log { \mu^2 \over 4 \Lambda^2}
+  V_{ct}(\phi_c)
 \nonumber \\
& & \quad \mbox{} + {T^4\over2\pi^2}\int_0^\infty x^2
\log\left( 1 - e^{-\rho(x)} \right) dx
\end{eqnarray}
in which the renormalization point $\mu$ is arbitrary
and $\rho(x)=\omega_k/T$ is the square root
\begin{equation}
\rho(x) = \sqrt{x^2 + V''(\phi_c)/T^2}.
\end{equation}
Thus if we choose the counterterms minimally
so that at $\phi_c$ they are
\begin{equation}
V_{ct}(\phi_c) = - {\Lambda^4 \over 16 \pi^2}
- {\Lambda^2 V''(\phi_c) \over 16 \pi^2}
+ {V''(\phi_c)^2 \over 64 \pi^2} \log { 4 \Lambda^2 \over \mu^2 },
\label{V_{ct}}
\end{equation}
then we obtain for the renormalized Helmholtz potential
the formula
\begin{eqnarray}
A_1(j,T) & = & V(\phi_c) + j\phi_c
+ {V''(\phi_c)^2 \over 64 \pi^2}
\left[\log{{V''(\phi_c) \over \mu^2}} + {1 \over 2} \right]
 \nonumber \\
& & \quad \mbox{} + {T^4\over2\pi^2}\int_0^\infty x^2
\log\left( 1 - e^{-\rho(x)} \right) dx,
\label{A_1(j,T) ren}
\end{eqnarray}
in which because of eq.(\ref{V_j stationary})
we may use either $\phi_c$ or $\bar \phi$ throughout.
The effective potential is then 
given by the Legendre transform (\ref{Vdef})
\begin{equation}
V_1(\phi_c,T) = A_1(j,T) - j \phi_c.
\label{V_1(phi_c,T) ren}
\end{equation}
\par
It will be useful in our discussion of 
the generalized effective potential
to develop further the relation
(\ref{phi_c is}) between the mean value $\phi_c$ 
and the minimum $\bar \phi$ of the altered potential $V_j(\phi)$.
By differentiating the extremal condition (\ref{dV_j = 0})
with respect to $j$,
we may find for the derivative of $\bar \phi$ the formula
\begin{equation}
{\partial \bar \phi \over \partial j}
= - \left[ {\partial^2 V(\bar \phi) \over \partial \bar \phi^2}
\right]^{-1}.
\label{dphi/dj}
\end{equation} 
It follows now from this formula and from
eqs.(\ref{phisdA}), (\ref{phi_c is}), and (\ref{A_1(j,T) ren})
that the mean value $\phi_c$ is to order $\hbar$
\begin{equation}
\phi_c = \bar \phi 
- { V'''(\bar \phi) \over 32 \pi^2}
\left[\log{{V''(\bar \phi) \over \mu^2}} + 1 \right]
- {T^2 V'''(\bar \phi) \over 4\pi^2 V''(\bar \phi)}
\int_0^\infty { x^2 dx \over 
\rho(x) \left( e^{\rho(x)} -1 \right)}
\label{phi_c = bar phi + etc}
\end{equation}
in which either $\phi_c$ or $\bar \phi$ may be used 
in the correction terms. 
\par
The archetypal example of a classical potential
that exhibits spontaneous symmetry breaking is
$V(\phi) = (\lambda/4) \left( \phi^2 - \sigma^2 \right)^2$. 
This potential has a positive second derivative
$V^{\prime\prime}(\phi) = \lambda \left( 3\phi^2 - \sigma^2 \right)$
only for fields $|\phi|$ that are greater than $\sigma / \sqrt{3}$.
For smaller $|\phi_c|$,
the one-loop effective potential $V_1(\phi_c,T)$ is complex.
According to eqs.(\ref{A_1(j,T) ren}) and (\ref{V_1(phi_c,T) ren}),
it is given by
\begin{eqnarray}
V_1(\phi_c,T) & = & {\lambda\over4}
(\phi_c^2-\sigma^2)^2
+ {\lambda^2 (3\phi_c^2 - \sigma^2 )^2 \over 64\pi^2}
\left[ \log\left({\lambda (3\phi_c^2 - \sigma^2 )
\over \mu^2}\right)
+ { 1 \over 2} \right] \nonumber \\
& & \quad \mbox{} + {T^4\over2\pi^2}\int_0^\infty x^2
\log\left( 1 - e^{-\rho(x)} \right) dx \label{olep}
\end{eqnarray}
where now $\rho(x)$ is
\begin{equation}
\rho(x) = \sqrt{x^2 + \lambda(3\phi_c^2-\sigma^2)/T^2}
\label{rho =}
\end{equation}
and $\mu$ is an arbitrary renormalization mass.
To this expression one may add arbitrary, finite
counterterms of the form
\begin{equation}
\lambda^2(A\phi_c^4+B\phi_c^2+C)
\label{ct's}
\end{equation}
from the renormalization of the hamiltonian $H$.
Due to the first logarithm 
$\log (\lambda (3\phi_c^2 - \sigma^2 )/\mu^2)$,
the effective potential $V_1(\phi_c,T)$
is complex for $|\phi_c| < \sigma/\sqrt{3}$,
where it is not possible to quantize the
approximate, altered theory.
\par
The effective potential is 
the sum of a function of $\phi_c^2 - \sigma^2$
and a function of $3\phi_c^2 - \sigma^2$,
and so cannot be convex. 
Its real part is concave, that is 
$\partial^2 \Re V_1(\phi_c,T)/\partial \phi_c^2 < 0$,
for most of the interval 
$-\sigma/\sqrt{3} < \phi_c < \sigma/\sqrt{3}$ 
for small $\lambda$, low temperatures $T$,
and reasonable renormalization. 
\par
For this example, the relation (\ref{phi_c = bar phi + etc})
between the mean value $\phi_c$ and the minimum $\bar \phi$ is
\begin{equation}
\phi_c = \bar \phi
- { 3 \lambda \bar \phi \over 16 \pi^2}
\left[\log{ \lambda(3\bar \phi^2-\sigma^2) \over \mu^2} + 1 \right]
- {3 T^2 \bar \phi \over 2\pi^2 (3\bar \phi^2-\sigma^2)}
\int_0^\infty { x^2 dx \over
\rho(x) \left( e^{\rho(x)} -1 \right)}
\label{phi_c = bar phi + eg}
\end{equation}
where $\rho(x)$ is given by eq.(\ref{rho =}).

\goodbreak

\section{\bf Where the Minima Lie}
\par
The multiple minima of symmetry-breaking
classical potentials typically divide
the space of fields into an inner region
that includes the point $\phi = 0$
and an outer region that extends to infinite fields.
The absolute minimum $\bar \phi$
of the altered potential $V_j(\phi) = V(\phi) + j \phi$
generally lies only in the outer region,
which is not the region of interest in studies 
of the early universe.
\par
Two examples may serve to illustrate this problem.
For the classical potential 
$V(\phi) = \lambda (\phi^2 - \sigma^2)^2/4$,
the current $j$ and the global minimum $\bar \phi$
of the altered potential 
$V_j(\phi) = \lambda (\phi^2 - \sigma^2)^2/4 + j \phi$ 
have opposite signs.
The absolute minimum $\bar \phi$ of 
$V_j(\phi)$ satisfies the extremal condition (\ref{dV_j = 0}) 
which we may write in the form
\begin{equation}
\bar \phi^2 = \sigma^2 - j / (\lambda \bar \phi)
\ge \sigma^2.
\end{equation}
Thus in this example, the global minimum $\bar \phi$, 
which is $\phi_c$ to lowest order in $\hbar$,
always lies in the region $|\bar \phi| \ge \sigma$.
\par
The second example is a theory
of four scalar fields with a classical potential 
$V(|\phi|) = a + b |\phi|^2 + c |\phi|^4$ 
that has as its the absolute minimum 
the hypersphere $|\phi|=\sigma$.
At the point $\phi_1=|\phi|$ and 
$\phi_2=\phi_3=\phi_4=0$, 
the matrix $V^{(2)}_{ij}(\phi)$ of second derivatives 
of the potential is
\begin{equation}
V^{(2)}_{ij} (\phi) = 
\mbox {\rm diag} 
\left(|\phi| V^{\prime\prime},V^\prime,V^\prime,V^\prime\right)
/|\phi|.
\end{equation}
Only in the outer region, $|\phi| > \sigma$, are
all the eigenvalues of this matrix positive.
Thus the absolute minimum $\bar \phi$ must lie in this exterior region.
\goodbreak

\section{\bf A More-Effective Potential}
\par
The reason for the complexity of the
effective potential in models exhibiting
spontaneous symmetry breaking is that 
the potential of the perturbed theory has the same 
second derivative $V_j^{\prime\prime}(\phi)$
as the original potential $V(\phi)$.
Somewhat in the spirit of the composite-operator 
technique~\cite{Corn 74, Hawk},
we may change the curvature of $V(\phi)$ by defining 
a more general Helmholtz potential $A(j,T;P)$
in which the linear probe $j \phi$ is replaced by
a quadratic polynomial $j P(\phi)$
\begin{equation}
e^{-\beta L^3 A(j,T;P)} =
\mbox{\rm Tr\,} e^{-\beta \left(H + \int\!\! j P(\phi) d^3x \right)}.
\label{def of MEA}
\end{equation}
On the one hand,
it is clear that by this device
we have not introduced any new divergences
into the theory.
On the other hand,
it is also clear that the polynomial $P(\phi)$ 
%which we have introduced as a perturbation
is itself singular and requires regularization.  
\par
Now the derivative of the Helmholtz potential $A(j,T;P)$
with respect to the external current $j$
is the mean value $P_c$
\begin{equation}
{\partial A(j,T;P) \over \partial j} = P_c
\label{dA = P_c}
\end{equation}
defined as the mean value $\langle P(\phi) \rangle$
of the quadratic form $P(\phi)$
\begin{equation}
P_c = {\mbox{\rm Tr\,} P(\phi(x))
e^{-\beta \left(H + \int\!\! j P(\phi) d^3x \right)} \over
\mbox{\rm Tr\,} e^{-\beta \left(H + \int\!\! j P(\phi) d^3x \right)}}.
\label{P_c =}
\end{equation}
The classical field $\phi_c$ is still the mean value
$\langle \phi \rangle$ of the quantum field $\phi$
\begin{equation}
\phi_c = {\mbox{\rm Tr\,} \phi(x)
e^{-\beta \left(H + \int\!\! j P(\phi) d^3x \right)} \over
\mbox{\rm Tr\,} e^{-\beta \left(H + \int\!\! j P(\phi) d^3x \right)}}.
\end{equation}
\par
We may now define a generalized effective potential
$V(\phi_c, T; P)$ as the Legendre transform 
of the Helmholtz potential $A(j,T;P)$,
\begin{eqnarray}
V(\phi_c, T; P) & = & 
A(j,T;P) - j {\partial A(j,T;P) \over \partial j}\nonumber \\
& = & A(j,T;P) - j P_c.
\label{def of MEV}
\end{eqnarray}
This notation may be confusing.
The variable that is conjugate to $j$ is $P_c$;
so strictly speaking we probably should write
$V(P_c, T;P)$ rather than $V(\phi_c, T;P)$.
But all the potentials considered in this paper
are actually and primarily functions of the external source $j$.
And for a given perturbative ground state,
the relationship between the source $j$ and the 
mean value $\phi_c$ is one to one.
Thus one may regard these potentials as functions of $\phi_c$, 
which is the physically more-significant variable.
\par
Like the conventional effective potential $V(\phi_c,T)$,
the generalized effective potential
at its minima
describes the unperturbed system.
For where the effective potential
$V(\phi_c, T; P)$
is stationary,
the external current $j$ vanishes
\begin{eqnarray}
0 & = & {\partial V(\phi_c,T; P) 
\over \partial \phi_c}\nonumber \\
& = & {\partial j \over \partial \phi_c}
\left( {\partial A(j,T;P) \over \partial j}
- P_c \right)
- j {\partial P_c \over \partial \phi_c} \nonumber \\
& = & - j {\partial P_c \over \partial \phi_c} 
\label{0 = j dP_c}
\end{eqnarray}
unless $P_c$ exceptionally should be independent
of $\phi_c$. 
\par
This more-effective potential $V(\phi_c, T; P)$ is real
but not necessarily convex.
Its second derivative contains two terms
\begin{equation}
{\partial^2 V(\phi_c,T; P) \over \partial \phi_c^2}
= - {\partial j \over \partial \phi_c} 
{\partial P_c \over \partial \phi_c}
- j {\partial^2 P_c \over \partial \phi_c^2} 
\end{equation}
and has no definite sign because
the first term is typically positive
while the second is typically negative.
\par
We shall need the relation between the mean value
$\phi_c$ and the external source $j$.
To that end one may introduce the 
further-generalized Helmholtz potential 
$A(j_1,j_2,T;P)$ defined by the relation
\begin{equation}
e^{-\beta L^3 A(j_1,j_2,T;P)} =
\mbox{\rm Tr\,}
e^{-\beta \left[ H 
+ \int \left( j_1 \phi + j_2 P \right) \, d^3x \right]}.
\label{A vg}
\end{equation} 
Clearly when $j_1$ vanishes,
this overly generalized Helmholtz potential
reduces to the generalized Helmholtz potential:
\begin{equation}
A(0,j_2,T;P) = A(j_2,T;P).
\label{too gen HP is gen HP}
\end{equation}
The mean value $\phi_c$
for the more-effective potential is thus
the partial derivative
\begin{equation}
\phi_c = \left. {\partial A(j_1,j_2,T;P)
\over \partial j_1 } \right|_{j_1 = 0}
\end{equation}
evaluated at $j_1 = 0$.
On the other hand,
the further-generalized Helmholtz potential
$A(j_1,j_2,T;P)$ is the usual Helmholtz potential $A(j_1,T)$
for a theory in which the classical potential $V(\phi)$ 
has been shifted to $V(\phi) + j_2 P(\phi)$  
\begin{equation}
A(j_1,j_2,T;P) = A(j_1,T)_{V+ j_2 P}.
\label{too gen HP is SUHP}
\end{equation}
Thus by differentiating that potential $A(j_1,T)_{V+ j_2 P}$ 
instead of $A(j_1,j_2,T;P)$
and by using eq.(\ref{phisdA}), we find
that for the more-effective potential
the mean value $\phi_c(j_2,T;P)$ is equal to 
the mean value $\phi_c(0,T)_{V+j_2P}$ associated with
the usual effective potential  
for a theory with a shifted potential $V+j_2P$
\begin{equation}
\phi_c(j_2,T;P) = \phi_c(0,T)_{V+ j_2 P}
\label{phic is SUphic}
\end{equation}
evaluated at $j_1 = 0$.
\par
By combining eqs.(\ref{too gen HP is gen HP}) and (\ref{too gen HP is SUHP}),
we find that the generalized Helmholtz potential $A(j_2,T;P)$
is the usual Helmholtz potential $A(0,T)_{V+ j_2 P}$ 
for the shifted potential at vanishing $j_1$
\begin{equation}
A(j_2,T;P) = A(0,T)_{V+ j_2 P}.
\label{gen HP is SUHP}
\end{equation}
\par
In supersymmetric theories it may be appropriate
to further generalize the perturbation $P$ to
a polynomial in both Fermi and Bose fields.
\goodbreak

\section{\bf The Meaning of Effective Potentials}
\par
The meaning of an effective potential is clearest at
zero temperature. 
Since the perturbed hamiltonian $H + \int j P(\phi) d^3x$
is hermitian, it has eigenstates $| j \rangle$
with energy $E_j$
\begin{equation}
\left( H + \int j P(\phi) d^3x \right)
| j \rangle = E_j | j \rangle .
\label{j state}
\end{equation}
Thus by its definition (\ref{def of MEA}),
the Helmholtz potential $A(j,T;P)$ 
in the limit $\beta \to \infty$ 
becomes the energy density 
$A(j,0;P) = E_j / L^3$
of the eigenstate $|j\rangle$ of the
altered hamiltonian $H + \int j P(\phi) d^3x$ 
with minimum energy $E_j$.
And so by eq.(\ref{def of MEV}), 
the effective potential 
\begin{equation}
V(\phi_c,0;P) = A(j,0;P) - j P_c 
\end{equation}
is the mean value of the hamiltonian density 
in this state $|j\rangle$
\begin{equation}V(\phi_c,0;P) = {\langle j|H|j\rangle \over L^3} . 
\label{Vmeans}
\end{equation}
And from eq.(\ref{0 = j dP_c}), it follows that 
the effective potential $V(\phi_c,0;P)$ 
at its absolute minimum 
is in general the energy density of the ground state $|0\rangle$
of the unperturbed theory, {\it i.e.,} the energy density
of the physical vacuum. 
\par
At finite temperatures,
the potential $A(j,T;P)$ is 
the Helmholtz free-energy density
of the mixture 
\begin{equation}
\rho(j) = {e^{-\beta \left[H + \int j P(\phi) d^3x\right] }
\over \mbox{\rm Tr\,}
e^{-\beta \left[H + \int j P(\phi) d^3x\right] }}
\label{mixture}
\end{equation}
associated with the altered hamiltonian density 
$\left[H + \int j P(\phi) d^3x\right]/L^3$.
The finite-temperature effective potential $V(\phi_c,T;P)$ is
the analog of the Gibbs free-energy density
\begin{equation}
V(\phi_c,T;P) = A(j,T;P) - j { \partial A(j,T;P) \over \partial j }
= A(j,T;P) - j P_c
\label{V and A}
\end{equation}
of this mixture.
By differentiating the definition (\ref{def of MEA})
of the Helmholtz free-energy density $A(j,T;P)$ 
with respect to the temperature $T$,
one may relate it to the perturbed energy density 
$u(j) = \mbox{\rm Tr\,} \left\{
\rho(j) \left[H + \int j P(\phi) d^3x\right] \right\} / L^3 $ and the 
entropy density $s = - \mbox{\rm Tr\,} \rho(j) \log \rho(j) / L^3 $
of the mixture (\ref{mixture}) by the equation
\begin{equation}
A(j,T;P) = u(j) - T s.
\end{equation}
It follows then from the relation (\ref{V and A}) 
that the more-effective potential or
Gibbs free-energy density $V(\phi_c,T;P)$
is related to
the energy density $u = \mbox{\rm Tr\,} \rho(j) H / L^3 $ 
and entropy density $s$ of the mixture $\rho(j)$
by the simpler equation 
\begin{equation}
V(\phi_c,T;P) = u - T s.
\end{equation}
Unfortunately the mixture $\rho(j)$ given by eq.(\ref{mixture})
coincides with the unperturbed physical mixture 
$\rho = e^{-\beta H}/ \mbox{\rm Tr\,} e^{-\beta H}$
only at the minima of $V(\phi_c,T;P)$
where the source $j$ vanishes.
\par
From this discussion it is clear that 
the choice of the polynomial $P(\phi)$
influences the effective potential $V(\phi_c,T;P)$ 
except at its various minima.
In particular if one uses a quadratic polynomial $P$
rather than a linear one, then one can avoid
spurious complexities. 
\goodbreak

\section{\bf The One-Loop More-Effective Potential}
\par
Although the usual effective potential $V(\phi_c,T)$
is satisfactory for theories in which the
classical potential $V(\phi)$ has positive
curvature, it can become complex when $V''(\phi) < 0$,
as in theories that exhibit
spontaneous symmetry breaking.
This section is concerned with the calculation
of the one-loop, more-effective potential $V(\phi_c,T;P)$
for an arbitrary renormalizable classical potential $V(\phi)$.
The computation will
closely follow that of the usual effective potential.
\par
If the curvature $V''(\phi)$ 
of the classical potential $V(\phi)$ 
is positive, 
then one may take $P(\phi) = \phi $
and the two effective potentials are identical.
If the curvature $V''(\phi)$ 
of the classical potential $V(\phi)$ 
is negative for some range of $ \phi $,
then the probe $P(\phi)$ should be quadratic.
There are then two cases according to whether
the classical potential has one or two minima.
\par
If the potential $V(\phi)$
has a single minimum which we may call $\phi_1$, 
then we may take the probe $P(\phi)$
to be half the square of the distance from $\phi_1$
\begin{equation}
P(\phi) \equiv P_1(\phi) = {1 \over 2 }
\left( \phi - \phi_1 \right) ^2 .
\label{def of P_1}
\end{equation}
\par
If the classical potential $V(\phi)$ has
two minima, which we may call $\phi_1$
and $\phi_2$, then it also has
one local maximum $\phi_0$ between them.
In this case we may take the probe $P(\phi)$
to be half the square of the distance from
the local maximum $\phi_0$
\begin{equation}
P(\phi) \equiv P_2(\phi) 
= {1\over 2} \left( \phi - \phi_0 \right)^2.
\label{def of P_2}
\end{equation}
\par
In both cases
the first step is to find the minima $\bar \phi$  
of the altered potential
$V_j = V(\phi) + j P(\phi)$. 
These minima $\bar \phi$ are roots of the equation
\begin{equation}
0 = V'(\bar \phi) + j P'(\bar \phi).
\label{bar phi vs j}
\end{equation}
Without any loss of generality,
we may let the leading term
of the potential $ V(\phi) $ be 
$ \lambda \phi^4 /4 $. 
\par
In the case of a unique minimum $\phi_1$,
the derivative
$ V'(\phi) $ then will be the product 
of the three factors
\begin{equation}
V'(\phi) = \lambda \left( \phi - \phi_1 \right)
\left( \phi - z \right) \left( \phi - z^\ast \right).
\label{V' =}
\end{equation}
So since the derivative of the probe $P_1(\phi)$ is
\begin{equation}
P'_1(\phi) = \left( \phi - \phi_1 \right),
\label{P'_1 = }
\end{equation}
the minima $\bar \phi$
of the altered potential $V_j(\phi)$ are given by
the quadratic equation
\begin{equation}
j = - { V'(\bar \phi) \over P'(\bar \phi)}
= - \lambda \left( \bar \phi - z \right) 
\left( \bar \phi - z^\ast \right) \le 0.
\label{j_1 is given by}
\end{equation}
We choose to compute the more-effective potential
about the root $\bar \phi$ that
is the absolute minimum of the altered potential
$V_j(\phi)$\null.
\par
When the potential 
$V(\phi)$ has two minima $\phi_1$ and $\phi_2$, 
its derivative $ V'(\phi) $ is the product 
of three factors
\begin{equation}
V'(\phi) = \lambda \left( \phi - \phi_0 \right)
\left( \phi - \phi_1 \right) \left( \phi - \phi_2 \right).
\label{V'_2 =}
\end{equation}
So since the derivative of the probe $P_2(\phi)$ is
\begin{equation}
P'_2(\phi) = \left( \phi - \phi_0 \right),
\label{P'_2 = }
\end{equation}
the minima $\bar \phi$
of the altered potential 
$V_j(\phi) = V(\phi) + j P_2(\phi)$
are given by the quadratic equation
\begin{equation}
j = - { V'(\bar \phi) \over P'_2(\bar \phi)}
= - \lambda \left( \bar \phi - \phi_1 \right) 
\left( \bar \phi - \phi_2 \right).
\label{j_2 is given by}
\end{equation}
The current $j$ is positive 
for $\phi_1 < \bar \phi < \phi_2$.
We choose to compute the more-effective potential
about the root $\bar \phi$ that
is the absolute minimum of the altered
potential $V(\phi)$\null.
\par
By its definition (\ref{def of MEA}),
the Helmoltz potential $A_0(j,T;P)$
in both cases and 
to lowest order in $\hbar$ is
\begin{equation}
A_0(j,T;P) = V(\bar \phi) + j P(\bar \phi).
\label{A_0g}
\end{equation}
\par
To find the order-$\hbar$ correction to this result,
we replace the altered potential $V+jP$
in the definition (\ref{def of MEA})
of the generalized Helmoltz potential $A(j,T;P)$
by its truncated Taylor series 
\begin{eqnarray}
V(\phi) + j P(\phi) & \approx & V(\bar \phi) + j P(\bar \phi)
\nonumber \\
& & \mbox{} + {1\over 2} \left( V''(\bar \phi) + j P''(\bar \phi)\right)
( \phi - \bar\phi )^2
\end{eqnarray}
and thereby reduce the problem to one that we have already solved
(\ref{A_1(j,T) =}).
Thus including counterterms, we find 
\begin{equation}
A_1(j,T;P) = V(\bar \phi) + V_{ct}(\bar \phi)
+ j \left( P(\bar \phi) + P_{ct}(\bar \phi) \right)
- {\log Z(\beta) \over \beta L^3}.
\end{equation}
Here $V_{ct}(\bar \phi)$ is the quartic polynomial (\ref{V_{ct}})
of counterterms that we used to renormalize
the ordinary effective potential;
$P_{ct}(\phi)$ is a quadratic
polynomial in $\phi$ that we shall use to regularize
the singular polynomial $P(\phi)$;
and $P_{ct}(\bar \phi)$ is that polynomial with the
field $\phi$ replaced by its mean value $\bar \phi$.
The squared mass in $Z(\beta)$ is positive
and writable simply as
\begin{equation}
m^2 = V''(\bar \phi) + j P''(\bar \phi) = V''(\bar \phi) + j \ge 0
\end{equation}
because $\bar \phi$ is a minimum of $V_j(\phi)$
and because $P''(\bar \phi)=1$ by construction
(\ref{def of P_1}--\ref{def of P_2}).
Thus the Helmholtz potential is
\begin{eqnarray}
A_1(j,T;P) & = & V(\bar \phi) + V_{ct}(\bar \phi) 
+ j \left( P(\bar \phi) + P_{ct}(\bar \phi) \right)
\nonumber \\
& & \quad \mbox{} + \int\!\! {d^3k \over (2\pi)^3}
\left[{\omega_k \over 2}
+ {\log\left(1 - e^{-\beta \omega_k }\right) \over \beta}
\right]
\label{A_1(j,T;P) =}
\end{eqnarray}
in which now
\begin{equation}
\omega_k = \sqrt{ k^2 + V^{\prime\prime}(\bar \phi) + j}.
\end{equation}
\par
If we again introduce a cut-off $\Lambda$
and perform the integration,
then we find
\begin{eqnarray}
A_1(j,T;P) & = & V(\bar \phi) + j P(\bar \phi)
+ { [ V''(\bar \phi)+ j ]^2 \over 64 \pi^2}
\left[\log{{V''(\bar \phi) + j \over \mu^2}} + {1 \over 2} \right]
 \nonumber \\
& & \mbox{}
+ {\Lambda^4 \over 16 \pi^2}
+ {\Lambda^2 [ V''(\bar \phi) + j ] \over 16 \pi^2}
+ {[ V''(\bar \phi) + j ]^2\over 64 \pi^2} \log { \mu^2 \over 4 \Lambda^2}
 \nonumber \\
& & \mbox{}
+  V_{ct}(\bar \phi) + j P_{ct}(\bar \phi)  
+ {T^4\over2\pi^2}\int_0^\infty x^2
\log\left( 1 - e^{\rho(x)} \right) dx
\end{eqnarray}
in which $\mu$ is the renormalization point and $\rho(x)$ is
the square root 
\begin{equation}
\rho(x) = \sqrt{x^2 + \left[ V''(\bar \phi) + j \right]/T^2}.
\label{rho(x)}
\end{equation}
Using the same counterterms $V_{ct}(\bar \phi)$
as the ones (\ref{V_{ct}}) we used for the usual effective potential,
we obtain
\begin{eqnarray}
A_1(j,T;P) & = & V(\bar \phi) + j P(\bar \phi)
+ { [ V''(\bar \phi)+ j ]^2 \over 64 \pi^2}
\left[\log{{V''(\bar \phi) + j \over \mu^2}} + {1 \over 2} \right]
 \nonumber \\
& & \mbox{}
+ {\Lambda^2 j \over 16 \pi^2}
+ { 2 j V''(\bar \phi) + j^2 \over 64 \pi^2}
\log { \mu^2 \over 4 \Lambda^2} + j P_{ct}(\bar \phi) 
 \nonumber \\
& & \mbox{} 
+ {T^4\over2\pi^2}\int_0^\infty x^2
\log\left( 1 - e^{\rho(x)} \right) dx.
\label{Coleman}
\end{eqnarray}
\par
All the terms on the right-hand side
of this equation for $ A_1(j,T;P) $
are functions of $j$,
$T$, and the parameters of the theory.
The same is true of $\bar \phi$.
In particular the minimal choice of
counterterms $ P_{ct}(\phi) $ 
evaluated at $\bar \phi$ is
\begin{equation}
P_{ct}(\bar \phi) = - {\Lambda^2 \over 16 \pi^2}
+ { 2 V''(\bar \phi) + j \over 64 \pi^2 }
\log { 4 \Lambda^2 \over \mu^2 }
\label{ct of j}
\end{equation}
when expressed as a mixed function of $j$,
$\bar \phi$, and the parameters of the theory.
We shall now use the relations
(\ref{j_1 is given by}) and (\ref{j_2 is given by})
that link $j$ and $\bar \phi$ to write 
the counterterms $ P_{ct}(\bar \phi) $
as functions of $\bar \phi$ 
and the parameters of the theory
without $j$.
The counterterms $ P_{ct}(\phi) $
will then be apparent.
\par
In the case in which the classical 
potential $V(\phi)$ possesses
a unique minimum $\phi_1$,  
we may identify the counterterms $ P_{1,ct}(\phi) $
associated with the quadratic probe $ P_1(\phi) $
by using the relation (\ref{j_1 is given by})
between $j$ and $\bar\phi$
to write the coefficient $j$ 
of the logarithmically divergent term in (\ref{ct of j})
as $ j = - \lambda \left| \bar \phi - z \right|^2$.
The minimal choice of counterterms $ P_{1,ct}(\phi) $
then is
\begin{equation}
P_{1,ct}(\phi) = - {\Lambda^2 \over 16 \pi^2}
+ { 2 V''(\phi) - \lambda \left| \phi - z \right|^2
\over 64 \pi^2 }
\log { 4 \Lambda^2 \over \mu^2 }.
\label{P_{ct}}
\end{equation}
These counterterms $P_{1,ct}(\phi)$
are of the same form as the probe $P_1(\phi)$,
to wit a quadratic polynomial in the variable $\phi$. 
\par
In the case in which the classical potential $V(\phi)$
possesses two minima $\phi_1$ and $\phi_2$,
we exploit the relationship (\ref{j_2 is given by})
between $j$ and $\bar\phi$
to write the same coefficient $j$
of the logarithmically divergent term in (\ref{ct of j}) as
$ j = - \lambda \left( \bar \phi - \phi_1 \right)
\left( \bar \phi - \phi_2 \right)$.
The minimal choice of counterterms $ P_{2,ct}(\phi) $
then is
\begin{equation}
P_{2,ct}(\phi) = - {\Lambda^2 \over 16 \pi^2}
+ { 2 V''(\phi) - \lambda \left( \phi - \phi_1 \right)
\left( \phi - \phi_2 \right) \over 64 \pi^2 }
\log { 4 \Lambda^2 \over \mu^2 }.
\label{P_{2,ct}}
\end{equation}
Like the probe $P_2(\phi)$,
these counterterms $P_{2,ct}(\phi)$
are a quadratic polynomial in the field $\phi$.
\par
The reason for the specific choices (\ref{def of P_1})
and (\ref{def of P_2}) of the probe polynomials
$ P_{1}(\phi) $ and $ P_{2}(\phi) $ 
was so that their counterterms
$ P_{1,ct}(\phi) $ and $ P_{2,ct}(\phi) $
would be of the same form as the probe polynomials themselves.
\par
With the counterterms $P_{ct}(\phi)$,
the Helmholtz potential is now
\begin{eqnarray}
A_1(j,T;P) & = & V(\bar \phi) + j P(\bar \phi)
+ { [ V''(\bar \phi)+ j ]^2 \over 64 \pi^2}
\left[\log{{V''(\bar \phi) + j \over \mu^2}} + {1 \over 2} \right]
\nonumber \\
& & \mbox{}
+ {T^4\over2\pi^2}\int_0^\infty x^2
\log\left( 1 - e^{\rho(x)} \right) dx.
\label{gen A_1 ren}
\end{eqnarray}
This formula and its counterpart (\ref{A_1(j,T) ren})
with $\phi_c$ replaced by $\bar \phi$
are an example of the relation (\ref{gen HP is SUHP})
between the generalized Helmholtz potential $A(j,T;P)$ and the
usual Helmholtz potential $A(j,T)_{V+jP}$ for the theory with
shifted potential $V + j P$. 
\par
To compute the more-effective potential $V_1(\bar \phi,T;P)$,
one must exploit the relationship (\ref{j_1 is given by})
or (\ref{j_2 is given by})
between $\bar \phi$ and $j$
and use eqs.(\ref{phi_c = bar phi + etc}) 
and (\ref{phic is SUphic}) to relate the mean
value $\phi_c$ to the minimum $\bar \phi$:
\begin{equation}
\phi_c = \bar \phi
- { V'''(\bar \phi) \over 32 \pi^2}
\left[\log{{V''(\bar \phi) + j \over \mu^2}} + 1 \right]
- {T^2 V'''(\bar \phi) \over 4\pi^2 \left( V''(\bar \phi) + j \right)}
\int_0^\infty { x^2 dx \over
\rho(x) \left( e^{\rho(x)} -1 \right)}
\label{gen phi_c = bar phi + etc}
\end{equation}
in which either $\phi_c$ or $\bar \phi$ may be used
in the correction terms and $\rho(x)$ is the square root
(\ref{rho(x)}).
By again using the relationship (\ref{j_1 is given by})
or (\ref{j_2 is given by})
between $\bar \phi$ and $j$,
one finally may perform the Legendre transform (\ref{def of MEV})
and compute the more-effective potential $V_1(\bar \phi,T;P)$.
We shall carry out this procedure explicitly
for the classical potential $V(\phi) = (\lambda/4)(\phi^2 - \sigma^2)^2$.
\goodbreak

\section{\bf An Example}
\par
This section contains a detailed
computation of the generalized
effective potential in the case
of the classical potential 
\begin{equation}
V(\phi) = {\lambda \over 4} \left( \phi^2 - \sigma^2 \right)^2. 
\label{V(phi)}
\end{equation}
We have seen in Sec.~V that
if we use the linear polynomial $P(\phi) = \phi$,
then the resulting effective potential 
is complex for $|\phi_c| < \sigma/3$. 
We shall find that by using the quadratic form
$P(\phi) = {\phi^2 / 2}$, 
we may explore the whole region $|\phi_c| \le \sigma$
with a more-effective potential $V(\phi_c,T;P)$ that remains real.
\par
Since the potential (\ref{V(phi)}) has indefinite
curvature and two minima,
the computation will follow the second of the
two cases discussed in Sec.~IX\null.
The minima are 
\begin{equation}
\phi_1 = - \sigma
\quad \hbox{\rm and} \quad \phi_2 = \sigma;
\label{minima}
\end{equation}
they are separated by a local maximum at
$\phi = \phi_0 = 0$\null.
So the probe is 
\begin{equation}
P(\phi) = {1 \over 2} \phi^2.
\label{P(phi)}
\end{equation}
The minima $\bar \phi$ of the altered potential
\begin{equation}
V_j(\phi) = V(\phi) +  {j \over 2} \phi^2 
\end{equation}
are the roots of the quadratic equation
\begin{equation}
j = - \lambda (\bar\phi^2 - \sigma^2)
\label{j=}
\end{equation}
or
\begin{equation}
\bar \phi^2 = \sigma^2 - j/\lambda. 
\end{equation}
Let us choose to quantize about the positive root 
\begin{equation}
\bar \phi = \bar \phi_+ = +\sqrt{\sigma^2 - (j/\lambda)}. 
\label{phi_c = f(j)}
\end{equation}
The mass associated with $\bar \phi$ is given by
\begin{equation}
m^2 = V_j^{\prime\prime}(\bar \phi) = 2\lambda\bar\phi^2\ge0.
\label{m^2}
\end{equation}
Since the current $j$ is related to the minimum $\bar \phi$ by 
(\ref{j=}),
we may write this squared mass also as
\begin{equation}
m^2 = 2\lambda\sigma^2 - 2j. 
\label{m^2 = m^2(j)}
\end{equation}
\par
To lowest order in $\hbar$, the Helmholtz potential $A_0(j,T;P)$ is
\begin{equation}
A_0(j,T;P) = V(\bar \phi) + j{\bar \phi^2\over2}
\label{A_0}
\end{equation}
in which $\bar \phi$ and $j$ are related by
$j = \lambda(\sigma^2 - \bar \phi^2)$.
%\label{j=f(phi)}
To this order the mean value $\phi_c$ and the minimum $\bar \phi$
are equal, and so
the more-effective potential $V_0(\phi_c,T;P)$ is
just the classical potential $V(\phi_c)$
\begin{equation}
V_0(\phi_c,T;P) = 
A_0(j,T;P) - j {\partial A(j,T;P) \over \partial j} 
= {\lambda\over4} (\phi_c^2-\sigma^2)^2.
\label{V_0=}
\end{equation}
\par
At the one-loop level, 
one finds by using eqs.(\ref{minima}) and (\ref{m^2})
that the regularizing counterterms (\ref{P_{2,ct}}) 
evaluated at $\bar \phi$ are 
\begin{equation}
P_{ct}(\bar \phi) = - {\Lambda^2 \over 16 \pi^2}
+ {\lambda (5\bar \phi^2 - \sigma^2) \over 64\pi^2}
\log{4 \Lambda^2 \over \mu^2}.
\end{equation}
As a function of the external source $j$,
the generalized Helmholtz potential $A_1(j,T;P)$ is then
given by eq.(\ref{gen A_1 ren}) as
\begin{eqnarray}
A_1(j,T;P) & = & {j\sigma^2 \over 2}
- {j^2 \over 4 \lambda}
+ {( 2\lambda \sigma^2 - 2j)^2 \over 64\pi^2}
\left[\log{2\lambda \sigma^2 - 2j \over \mu^2 } + {1 \over 2}
\right]
\nonumber \\
& & \mbox{}
+ {T^4\over2\pi^2}\int_0^\infty x^2
\log\left( 1 - e^{\rho(x)} \right) dx
\end{eqnarray}
where now $\rho(x)$ is the square root
\begin{equation}
\rho(x) = \sqrt{x^2 + (2\lambda \sigma^2 -2j)/T^2} 
= \sqrt{x^2 + 2\lambda \bar \phi^2/T^2}.
\end{equation}
\par
The generalized effective potential
is defined (\ref{def of MEV}) as the Legendre transform 
\begin{equation}
V_1(\phi_c, T; P) = A(j,T;P) - j {\partial A(j,T;P) \over \partial j}.
\end{equation}
By performing the indicated differentiation with respect to $j$
and by then expressing $j$ as $\lambda(\sigma^2-\bar \phi^2)$, 
we may write $V_1(\phi_c,T;P)$ as 
%in terms of the minimum $\bar \phi$ as
\begin{eqnarray}
V_1(\phi_c,T;P) & = & {\lambda \over 4} (\bar \phi^2 - \sigma^2)^2
+ {\lambda^2 \bar \phi^2 \over 32\pi^2}
\left[ (4 \sigma^2 - 2 \bar \phi^2) \log{2\lambda\bar \phi^2 \over \mu^2}
+ 4 \sigma^2 - 3 \bar \phi^2 \right] 
\nonumber \\
& & \mbox{}
+ {T^4\over2\pi^2}\int_0^\infty \!\! x^2 \left[
\log\left( 1 - e^{-\rho(x)} \right)
+{\lambda(\sigma^2 - \bar \phi^2)
\over T^2 \rho(x) \left( e^{\rho(x)} - 1 \right) } \right] \! dx.
\label{MEVT =}
\end{eqnarray}
In the correction terms, which are of order $\hbar$,
we may write indifferently $\bar \phi$ or $\phi_c$.
But in the first term $V(\bar \phi)$
we must use eq.(\ref{gen phi_c = bar phi + etc}) 
to distinguish $\phi_c$ 
\begin{equation}
\phi_c = \bar \phi - {3 \lambda \bar \phi \over 16 \pi^2}
\left[ \log \left( { 2 \lambda \bar \phi^2 \over \mu^2 } \right)
+ 1 \right]
- { 3 T^2 \over 4 \pi^2 \bar \phi }
\int_0^\infty \!\! {x^2 dx \over \rho(x) \left( e^{\rho(x)} - 1 \right) }
\end{equation}
from $\bar \phi$.
Since $\phi_c$ and $\bar \phi$ differ by terms of order $\hbar$,
we need keep only the leading term
\begin{equation}
V(\bar \phi) \approx V(\phi_c) + \lambda 
\phi_c \left( \phi_c^2 - \sigma^2 \right) (\bar \phi - \phi_c)
\end{equation}
and may switch now to the variable $\phi_c$ throughout
in the resulting formula for the one-loop,
finite-temperature, generalized effective potential:
\begin{eqnarray}
V_1(\phi_c,T;P) & = & {\lambda \over 4} (\phi_c^2 - \sigma^2)^2
+ {\lambda^2 \phi_c^2 \over 32\pi^2}
\left[ (4 \phi_c^2 - 2 \sigma^2) \log{2\lambda\phi_c^2 \over \mu^2}
+ 3 \phi_c^2 - 2 \sigma^2 \right]
\nonumber \\
& & \mbox{}
+ {T^4\over2\pi^2}\int_0^\infty \!\! x^2 \left[
\log\left( 1 - e^{-\rho(x)} \right)
+{\lambda(\phi_c^2 - \sigma^2)
\over 2 T^2 \rho(x) \left( e^{\rho(x)} - 1 \right) } \right] \! dx
\quad
\label{MEVT = hot stuff}
\end{eqnarray}
in which $\rho(x)$ is the square root
\begin{equation}
\rho(x) = \sqrt{x^2 + 2\lambda \phi_c^2 / T^2 }.
\label{row}
\end{equation}
\par
One may adopt a specific set of renormalization
conditions by adding finite counterterms
to the preceding formula.
A sensible set of conditions is $V_1(\sigma,0;P) = 0$,
$V'_1(\sigma,0;P) = 0$, 
and $V''_1(\sigma,0;P) = m_H^2 = 2\lambda \sigma^2$.
We may satisfy these conditions by adding 
the quartic polynomial
\begin{equation}
{\lambda^2 \over 32 \pi^2} \left( -8\phi^4 
+ 10 \sigma^2 \phi^2 -3 \sigma^4 \right)
\label{mep ct's}
\end{equation}
to the more-effective potential (\ref{MEVT = hot stuff})
and setting $\mu^2 = 2\lambda \sigma^2$.
The resulting expression is
\begin{eqnarray}
V_1(\phi_c,T;P) & = & {\lambda \over 4} (\phi_c^2 - \sigma^2)^2
\nonumber \\
& & \mbox{}
+ {\lambda^2 \over 32\pi^2}
\left[ (4 \phi_c^4 - 2 \sigma^2 \phi_c^2) 
\log{\phi_c^2 \over \sigma^2}
- 5 \phi_c^4 + 8 \sigma^2 \phi_c^2 - 3 \sigma^4 \right]
\quad
\nonumber \\
& & \mbox{}
+ {T^4\over2\pi^2}\int_0^\infty \!\!\! x^2 \! \left[
\log\left( 1 - e^{-\rho(x)} \right)
+{\lambda(\phi_c^2 - \sigma^2)
\over 2 T^2 \rho(x) \left( e^{\rho(x)} - 1 \right) } \right] \! dx
\quad
\label{MEVT AH}
\end{eqnarray}
with $\rho(x)$ given by eq.(\ref{row}).
If we set $\mu^2 = 2\lambda \sigma^2$
in the usual effective potential (\ref{olep})
and add to it these same counterterms, 
then it becomes
\begin{eqnarray}
V_1(\phi_c,T) & = & {\lambda\over4}
(\phi_c^2-\sigma^2)^2 \nonumber \\
& & \mbox{} + {\lambda^2 \over 64\pi^2}
\left[ (3\phi_c^2 - \sigma^2 )^2 
\log\left({ 3\phi_c^2 - \sigma^2  \over 2 \sigma^2}\right)
- { 23 \over 2} \phi_c^4 + 17 \sigma^2 \phi_c^2 
- { 11 \over 2} \sigma^4 \right] \nonumber \\
& & \mbox{} + {T^4\over2\pi^2}\int_0^\infty x^2
\log\left( 1 - e^{-\rho(x)} \right) dx 
\label{aholep}
\end{eqnarray}
where now $\rho(x)$ is
\begin{equation}
\rho(x) = \sqrt{x^2 + \lambda(3\phi_c^2-\sigma^2)/T^2}.
\label{ahrho =}
\end{equation}
\par
A numerical analysis of the formula (\ref{MEVT AH})
shows that the critical temperature 
runs from $T_C \approx 1.81 \, \sigma$ for $\lambda = 0.1$
to $T_C \approx 1.7413 \, \sigma$ for $\lambda = 1$.
At higher temperatures, the absolute minimum of
$V_1(\phi_c,T;P)$ is at $\phi_c = 0$; at lower temperatures
it is at $\phi_c > 0.62 \, \sigma$ for $\lambda = 0.1$
and at $\phi_c > 0.69 \, \sigma$ for $\lambda = 1$.
The transition is weakly first order because at $T = T_C$
the barrier separating the two minima is slight.
At the barrier temperature, which shifts 
from $T_B \approx 1.87 \, \sigma$
for $\lambda = 0.1$ to $T_B \approx 1.865 \, \sigma$
for $\lambda = 1$,
this barrier disappears, and the field $\phi_c$ can roll
classically from $\phi_c = \sigma$ to the absolute
minimum at $\phi = 0$.
\par
By differentiating eq.(\ref{MEVT AH})
with respect to $\phi_c$ at $\phi_c = 0$,
one may show that
at all positive temperatures, 
the derivative of the more-effective potential
at $\phi_c = 0+\epsilon$ is positive
\begin{equation}
\left.{\partial V_1(\phi_c,T;P) \over \partial \phi_c}
\right|_{\phi_c=0+\epsilon} > 0.
\end{equation}
Thus the point $\phi_c = 0$ is a local minimum at all $T > 0$.
The temperature $T_2$ at which this minimum
disappears is therefore zero. 
In models of the early universe,
inflation can occur if the field $\phi_c$ sticks
at this local minimum.
\par
The formula (\ref{aholep})
for the usual effective potential $V_1(\phi_c,T)$
for $\lambda = 1$
is plotted (solid lines) in Fig.~1 
at various temperatures.
The region $0 \leq \phi_c < \sigma/\sqrt{3}$
is blank because the usual effective potential $V_1(\phi_c,T)$
is complex in this region.
Clearly it is not possible to determine
the critical temperature $T_C$ or the barrier
temperature $T_B$ from the standard effective potential. 
The dotted curve represents the classical potential $V(\phi)$.
The vertical axes are in units of $V(0)=\lambda\sigma^4/4$.
\par
The formula (\ref{MEVT AH}) for 
the generalized effective potential $V_1(\phi_c,T;P)$
is plotted (solid lines) in Fig.~2
for $\lambda = 1$
at various temperatures including
the critical temperature $T_C \approx 1.7413 \, \sigma$,
at which the minima at $\phi_c = 0$ and $\phi_c \approx 0.7 \, \sigma$
are equally deep, and the barrier temperature 
$T_B \approx 1.865 \, \sigma$,
at which the barrier between these minima disappears.
As in Fig.~1, the dotted curve represents 
the classical potential $V(\phi)$.
\par
Figure~3 displays 
the generalized effective potential 
$V_1(\phi_c,T;P)$ as solid lines and 
the usual effective potential $V_1(\phi_c,T)$ as dashed lines
for the same classical potential $V(\phi)$  (dots)
with $\lambda = 1$.
The more-effective potential tracks
the standard effective potential so closely  
where the latter is real ($\phi_c \geq \sigma/\sqrt{3}$)
that the dashes of $V_1(\phi_c,T)$ are visible only
near $\phi_c = \sigma/\sqrt{3}$ and near $\phi_c = 1.4 \, \sigma$.
%This close agreement precluded the use of a single figure.
The generalized effective potential
naturally extends the usual effective potential to $\phi_c = 0$.
\par
By using the Haber--Weldon expansions of
Bose--Einstein integrals~\cite{Howie},
one may develop a high-temperature expansion~\cite{Tom and me}
for the generalized effective potential (\ref{MEVT AH})
\begin{eqnarray}
V_1(\phi_c,T;P) & = &
- {\pi^2 \over 90} T^4 
+ {\lambda \over 24} \left( 3 \phi_c^2 - \sigma^2 \right) T^2
- {\lambda^{3/2} \over 12 \sqrt{2} \, \pi} 
\left( 7|\phi_c|^3 - 3\sigma^2 |\phi_c| \right) T 
\nonumber \\
& & \mbox{} + {\lambda^2 \over 16 \pi^2} 
\left( 2 \phi_c^4 - \sigma^2 \phi_c^2 \right)
\log {T^2 \over \sigma^2} 
+ {\lambda \over 4 } \left( \phi_c^2 - \sigma^2 \right)^2
\nonumber \\
& & \mbox{} + {\lambda^2 \over 32 \pi^2}
\left( f\phi_c^4 + s\phi_c^2 - 3 \sigma^4 \right) 
+ {\cal O}({\sigma^2 \over T^2})
\label{Tom and me}
\end{eqnarray}
in which $f = 4 \log(8\pi^2/\lambda) - 8 \gamma$,
$s = 10 - 2 \log(8\pi^2/\lambda) - 4 \gamma$,
and $\gamma \approx 0.577 215 66$.
The high-temperature expansion~\cite{Dola 74}
of the usual effective potential
has the form
\begin{eqnarray}
V_1(\phi_c,T) & = &  - {\pi^2 \over 90} T^4 
+ {\lambda \over 24} \left( 3\phi_c^2 - \sigma^2 \right) T^2
- { \lambda^{3/2} \over 12 \pi }
\left( 3 \phi_c^2 - \sigma^2 \right)^{3/2} T 
\nonumber \\
& & \mbox{}
+ { \lambda^2 \over 64 \pi^2}
\left( 3 \phi_c^2 - \sigma^2 \right)^2 
\log { T^2 \over
 \lambda^2 \left( 3 \phi_c^2 - \sigma^2 \right)^2} 
+ {\cal O} (1).
\label{dolan}
\end{eqnarray}
\par
Although the third and fourth terms of this formula are
imaginary for $|\phi_c| < \sigma/\sqrt{3}$,
the first two terms, which are of order $T^4$
and $T^2$, do agree with the first two terms
of the high-temperature expansion (\ref{Tom and me})
of the generalized effective potential (\ref{MEVT AH}).  
Thus the seemingly outrageous approximation, 
often made in work on the early universe, 
of keeping only these two, real terms 
of the high-temperature expansion (\ref{aholep}) 
turns out to be a reasonable one after all.
\goodbreak
%\pagebreak

\section*{\bf  Acknowledgements}
\par
I am grateful to D.~Bailin, H.~Barnum, S.~Coleman,
B.~Grinstein, I.~Hinchliffe, J.~Jers\'{a}k,
S.~Johnson, G.~Keaton, E.~Kolb, H.~Leutwyler, S.~Mandelstam,
R.~Matzner, J.~March-Russell,
F.~Paige, R.~Reeder, H.~Stapp, G.~Stephenson, P.~Stevenson,
T.~Weiler, E.~Weinberg, R.~Xiu, and H.~Zhang
for helpful discussions.
This research was supported by the U.S. Department of Energy
under contracts DE-AC03-76SF00098 and DE-FG04-84ER40166.
Some of this work was done at the Aspen Center for Physics.
\goodbreak

\vfill \eject

\begin{thebibliography}{99}
  \bibitem{Heis} W.~Heisenberg and H.~Euler,
  {\sl Z.~Phys.\/} 98 (1936) 714.
  \bibitem{Schw} J.~Schwinger,
  {\sl Proc.~Natl.~Acad.~Sci.~USA\/} 37 (1951) 452 and
  {\sl Phys.~Rev.\/} 82 (1951) 664.
  \bibitem{Gold 62} J.~Goldstone, A.~Salam, and S.~Weinberg,
  {\sl Phys.~Rev.\/} 127 (1962) 965.
  \bibitem{Jona 64} G.~Jona-Lasinio,
  {\sl Nuovo Cimento\/} 34 (1964) 1790.
  \bibitem{Gold 61} J.~Goldstone, 
  {\sl Nuovo Cimento\/} 19 (1961) 15;
  Y.~Nambu and G.~Jona-Lasinio,
  {\sl Phys.~Rev.\/} 122 (1961) 345;
  124 (1961) 246.
  \bibitem{Cole 73} S.~Coleman and E.~Weinberg,
  {\sl Phys.~Rev.\/} D7 (1973) 1888;
  S.~Coleman, {\it Aspects of Symmetry\/}
  (Camb.\ Univ.\ Press, 1985), p.~132.
  \bibitem{Lind 76} A.~D.\ Linde,
  {\sl JETP Letters\/} 23 (1976) 64.
  \bibitem{Wein 76} S.~Weinberg,
  {\sl Phys.\ Rev.\ Lett.\/} 36 (1976) 294.
  \bibitem{West 76} P.~West,
  {\it Introduction to Supersymmetry and Supergravity\/}
  (World Scientific, 1990).
  \bibitem{Lind 72} D.~Kirzhnits and A.~D.\ Linde,
  {\sl Phys.\ Lett.\ B\/} 42 (1972) 471.
  \bibitem{KLind 74} D.~Kirzhnits and A.~D.\ Linde,
  {\sl JETP\/} 40 (1974) 628 and {\sl Ann.\ Phys.\/} 101 (1976) 195.
  \bibitem{Dola 74} L.~Dolan and R.~Jackiw,
  {\sl Phys.~Rev.\ D\/} 9 (1974) 3320.
  \bibitem{Wein 74} S.~Weinberg,
  {\sl Phys.~Rev.\ D\/} 9 (1974) 3357.
  \bibitem{Bern 74} C.~Bernard,
  {\sl Phys.~Rev.\ D\/} 9 (1974) 3312.
  \bibitem{Kolb 90} E.~W. Kolb and M.~S. Turner,
  {\it The Early Universe\/} (Addison-Wesley, 1990).
  \bibitem{loopsfail} Y.~Fujimoto, L.~O'Raifeartaigh, and
  G.~Parravicini, {\sl Nucl.\ Phys.\ B\/} 212 (1983) 268;
  R.~Haymaker and J.~Perez-Mercader,
  {\sl Phys.~Rev.\ D\/} 15 (1983) 1948;
   R.~J. Rivers, {\sl Z.~Phys.~C\/} 22 (1984) 137;
   {\sl Path-Integral Methods in Quantum Field Theory\/}
   (Cambridge, 1987), pp.\ 235--272;
   C.~M. Bender and F.~Cooper, {\sl Nucl.\ Phys.~B\/} 224 (1983) 403;
   F.~Cooper and B.~Freedman, {\sl Nucl.\ Phys.~B\/} 239 (1984) 459.
  \bibitem{real&convex}  K.~Symanzik, {\sl Commun.\ Math.\ Phys.\ \/}
   16 (1970) 48;
   J.~Iliopoulos, C.~Itzykson, and A.~Martin,
  {\sl Rev.\ Mod.\ Phys.\ \/} 47 (1975) 165.
  \bibitem{EW} E.~J. Weinberg and A.~Wu, {\sl Phys.~Rev.\ D\/}
  36 (1987) 2474.
  \bibitem{Chang} S.~J. Chang, {\sl Phys.~Rev.\ D\/} 12 (1975) 1071.
  \bibitem{Barnes} T.~Barnes and G.~I. Ghandour,
  {\sl Phys.~Rev.\ D\/} 22 (1980) 924.
  \bibitem{Stevenson} P.~M. Stevenson, {\sl Phys.~Rev.\ D\/} 
  30 (1984) 1712; 32 (1985) 1389; {\sl Z.~Phys.\/} C35 (1987) 467.
  \bibitem{Fukuda} R.\ Fukuda and E.\ Kyriakopoulos,
  {\sl Nucl.\ Phys.~B\/} 85~(1975)~354;
  R.\ Fukuda, {\sl Prog.\ Theor.\ Phys.\/} 56 (1976) 258.
  \bibitem{O'Raifeartaigh} L.\ O'Raifeartaigh, A.\ Wipf, and H.\ Yoneyama,
  {\sl Nucl.\ Phys.\ B\/} 271 (1986) 653.
  \bibitem{Ringwald} A.~Ringwald and C.~Wetterich,
  {\sl Nucl.\ Phys.\/} {B334} (1990) 506.
  \bibitem{Corn 74} J.~M. Cornwall, R.~Jackiw, and E.~Tomboulis,
  {\sl Phys.~Rev.\ D\/} 10 (1974) 2428.
  \bibitem{Hawk} S.~W. Hawking and I.~G. Moss,
  {\sl Nucl.\ Phys.~B\/} 224~(1983)~180.
  \bibitem{Lawrie} I.~D. Lawrie,
  {\sl Nucl.\ Phys.\/} {B301} (1988) 685.
  %\bibitem{Claude 80} C.~Itzykson and J.-M. Drouffe,
  %{\it Statistical Field Theory,} Vol.~1
  %(Cambridge University Press, 1989), p.~241. 
  \bibitem{Hall} G.~W. Anderson and L.~J. Hall,
  {Phys.\ Rev.\/} D45 (1992) 2685.
  \bibitem{Tom and me} K.~Cahill and T.~Weiler,
   work in progress.
  \bibitem{Howie} H.~E. Haber and H.~A. Weldon,
  {\sl J.~Math.~Phys.\/} 23~(1982)~1852.
\end{thebibliography}
\end{document}